\documentclass[9pt,conference]{IEEEtran}
\usepackage{dcase2026}
\usepackage{stfloats}
\usepackage{capt-of}
\newsavebox{\saidoverview}

\title{SAID: Semantic Acoustic Imaging Detector for Sound Event Localization and Detection}
\name{Runbang Wang$^{1,2,4}$, Zining Liang$^{2,4}$, Yin Cao$^{3}$, Qiuqiang Kong$^{2,4}$\sthanks{Corresponding author.}}
\address{$^{1}$Nanjing University, China\\
$^{2}$The Chinese University of Hong Kong, Hong Kong SAR, China\\
$^{3}$Institute of Acoustics, Chinese Academy of Sciences, China\\
$^{4}$Shun Hing Institute of Advanced Engineering (SHIAE), Hong Kong SAR, China\\
$^{1,2,4}$runbang@link.cuhk.edu.hk, $^{2,4}$violetliang@link.cuhk.edu.hk, $^{3}$yin.k.cao@gmail.com, $^{2,4}$qqkong@ee.cuhk.edu.hk}

\begin{document}
\setcounter{footnote}{1}
\maketitle
\begin{abstract}
In daily life, people hear speech, footsteps, and music around them.
We can often recognize these sounds and judge where they come from.
Each sound source can be shown on a separate acoustic map, a rectangular image covering $360^{\circ}$ horizontally and $180^{\circ}$ vertically.
The map shows the directions occupied by the source as a region and the sound energy within that region.
A class label identifies the sound.
Predicting these labeled acoustic maps from audio is called semantic acoustic imaging.
Such maps could help robots perceive their surroundings and allow augmented reality displays to show sound regions and classes over the real world.
Existing models can recognize sound classes and estimate a direction for each source.
However, a direction alone does not describe the source region or its energy.
Acoustic imaging must also distinguish sound sources in nearby directions, while the number of active sources and the regions they occupy can change over time.
We therefore propose the Semantic Acoustic Imaging Detector (SAID), which predicts a separate labeled acoustic map for each active source from audio.
First, we pretrain Audio2Sph, SAID's audio encoder, through sound energy estimation across directions without class labels.
Then, we train the complete SAID model to predict source regions, energy, and classes together.
We also develop a pipeline that generates simulated recordings for pretraining and supports fine-tuning on real recordings.
On the official DCASE2026 Task 3 Track A evaluation set, our submitted system ranks first with 0.1080 macro-averaged mean average precision (Macro mAP) and 0.3962 Macro Pearson $r$.
Demos and code are provided at \pdfstartlink attr {/Border [0 0 0]} user {/Subtype /Link /A << /S /URI /URI (https://github.com/IN03X/SAID) >>}\url{https://github.com/IN03X/SAID}\pdfendlink.
\end{abstract}

\section{Introduction}
\label{sec:introduction}

Speech, footsteps, and music are familiar sounds around us.
We can often recognize these sounds and tell where they come from.
Each sound source occupies a region that can be marked on a separate acoustic map of directions around the listener.
This map is a rectangular image covering $360^{\circ}$ horizontally and $180^{\circ}$ vertically.
It shows the sound energy within that region, and a class label such as speech or footsteps identifies the sound.
Estimating these labeled acoustic maps from audio is the task of semantic acoustic imaging~\cite{dcase2026task3}.
Robots could use this information to better understand their surroundings through sound.
Augmented reality could display acoustic maps and class labels over the user's view of the real world.

Semantic acoustic imaging builds on earlier research asking: which direction does a sound come from?
A basic starting point is to treat each sound source as a point and consider only sound arriving directly, without echoes~\cite{evers2020locata}.
Classical direction estimators include generalized cross-correlation with phase transform (GCC-PHAT)~\cite{knapp1976gcc} and multiple signal classification (MUSIC)~\cite{schmidt1986music}.
Later, the Acoustic Source Localization and Tracking (LOCATA) challenge tested localization in real rooms, where echoes could suggest incorrect directions~\cite{evers2020locata}.
Methods improved direction estimates and followed moving sources, without identifying classes~\cite{nakadai2018music,li2018tracking}.
The 2019 Detection and Classification of Acoustic Scenes and Events (DCASE) challenge then introduced sound event localization and detection (SELD)~\cite{politis2021dcase2019} in simulated recordings~\cite{adavanne2019multiroom}.
Models such as SELDnet jointly predicted class and direction~\cite{adavanne2018seldnet,cao2019twostage}.
Subsequently, DCASE2022~\cite{politis2022starss22} and DCASE2023~\cite{shimada2023starss23} brought SELD to real scenes with overlapping sounds.
Models evolved from convolutional recurrent~\cite{adavanne2018seldnet} to attention-based networks~\cite{park2021m2mast,hu2022realscenes}, with outputs distinguishing simultaneous same-class sources~\cite{shimada2022multiaccdoa}.
More recently, DCASE2026 replaced each sound source's direction with a region on an acoustic map~\cite{dcase2026task3}.
Image segmentation in computer vision also predicts regions and classes for objects such as people or cars in photographs~\cite{he2017mask}.
The DCASE2026 baseline adapts a mask region-based convolutional neural network (Mask R-CNN) for acoustic imaging~\cite{dcase2026baseline}.
MaskFormer~\cite{cheng2021maskformer} and Mask2Former~\cite{cheng2022mask2former} instead predict labeled regions through mask decoders.

Semantic acoustic imaging presents three challenges.
First, conventional SELD methods predict classes and directions, but not source regions or the energy within them~\cite{adavanne2018seldnet,shimada2022multiaccdoa}.
Second, the model must predict separate labeled acoustic maps for nearby sources whose number, positions, and region sizes change over time~\cite{dcase2026task3}.
Third, image segmentation can handle varying numbers of regions~\cite{cheng2022mask2former}, but audio recordings do not directly provide features arranged like an acoustic map~\cite{dcase2026baseline}.
To address these three challenges, we need features organized by direction to predict a separate labeled acoustic map for each active source.

We introduce the Semantic Acoustic Imaging Detector (SAID).
Audio2Sph, SAID's encoder, converts audio into panoramic features organized by direction, following the layout of an acoustic map.
Sph2Imaging, SAID's decoder, converts these features into labeled acoustic maps.
Our contributions are threefold.
First, we introduce Audio2Sph, an encoder pretrained through sound energy estimation across directions without class labels.
Second, we propose SAID, a model that predicts a separate labeled acoustic map for each active sound source from audio.
Third, we develop a data generation and training pipeline that generates simulated recordings during pretraining and supports fine-tuning on DCASE recordings.

\begin{lrbox}{\saidoverview}
\begin{minipage}{\columnwidth}
\centering
\includegraphics[page=4,trim=0 326bp 244bp 0,clip,width=\columnwidth]{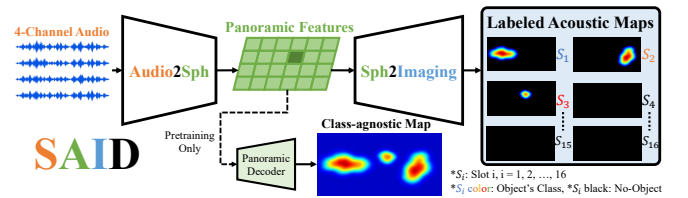}
\captionof{figure}{SAID overview. Audio2Sph encodes audio into panoramic features; Sph2Imaging predicts labeled acoustic maps. The dashed branch is used only for class-agnostic pretraining. Visualizations are schematic.}
\label{fig:said}
\end{minipage}
\end{lrbox}

\begin{figure*}[t]
\centering
\includegraphics[page=2,trim=0 173bp 0 0,clip,width=\textwidth]{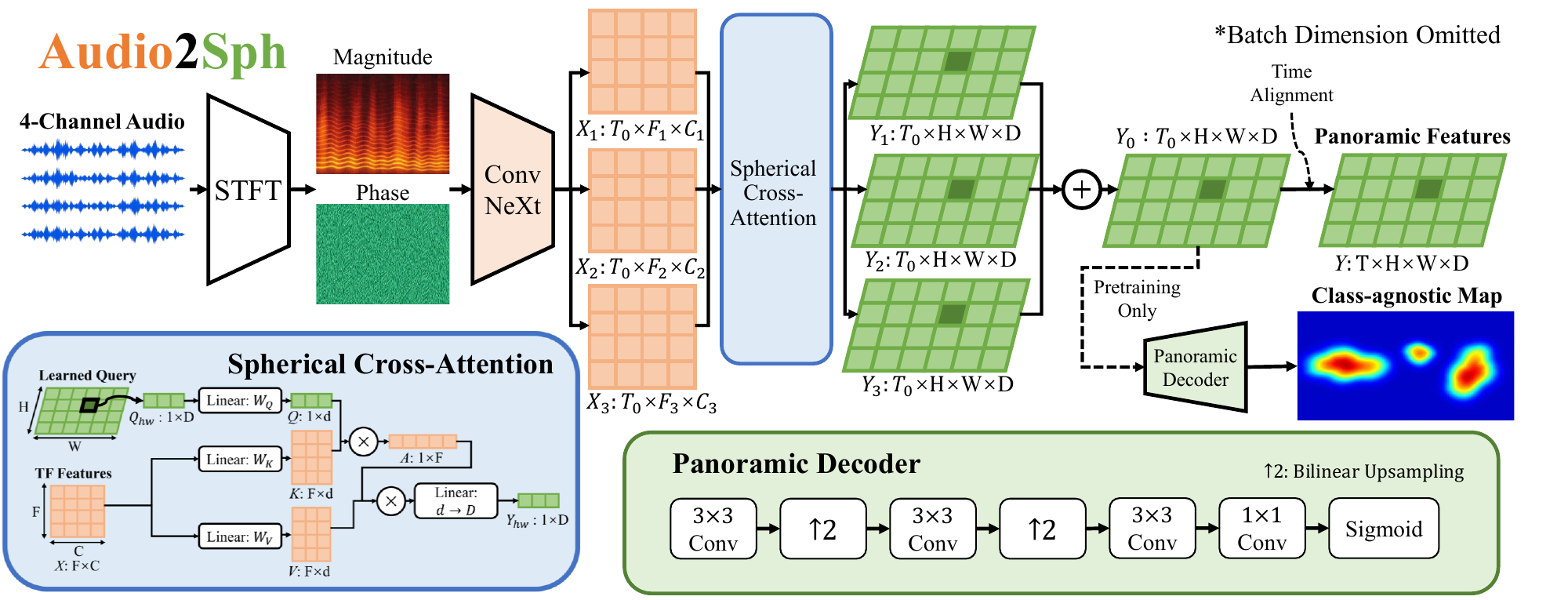}
\caption{Audio2Sph maps time--frequency features onto a directional grid through learned queries. The panoramic decoder predicts a class-agnostic map only during pretraining.}
\label{fig:audio2sph}
\end{figure*}

\begin{figure*}[t]
\centering
\includegraphics[page=3,width=\textwidth]{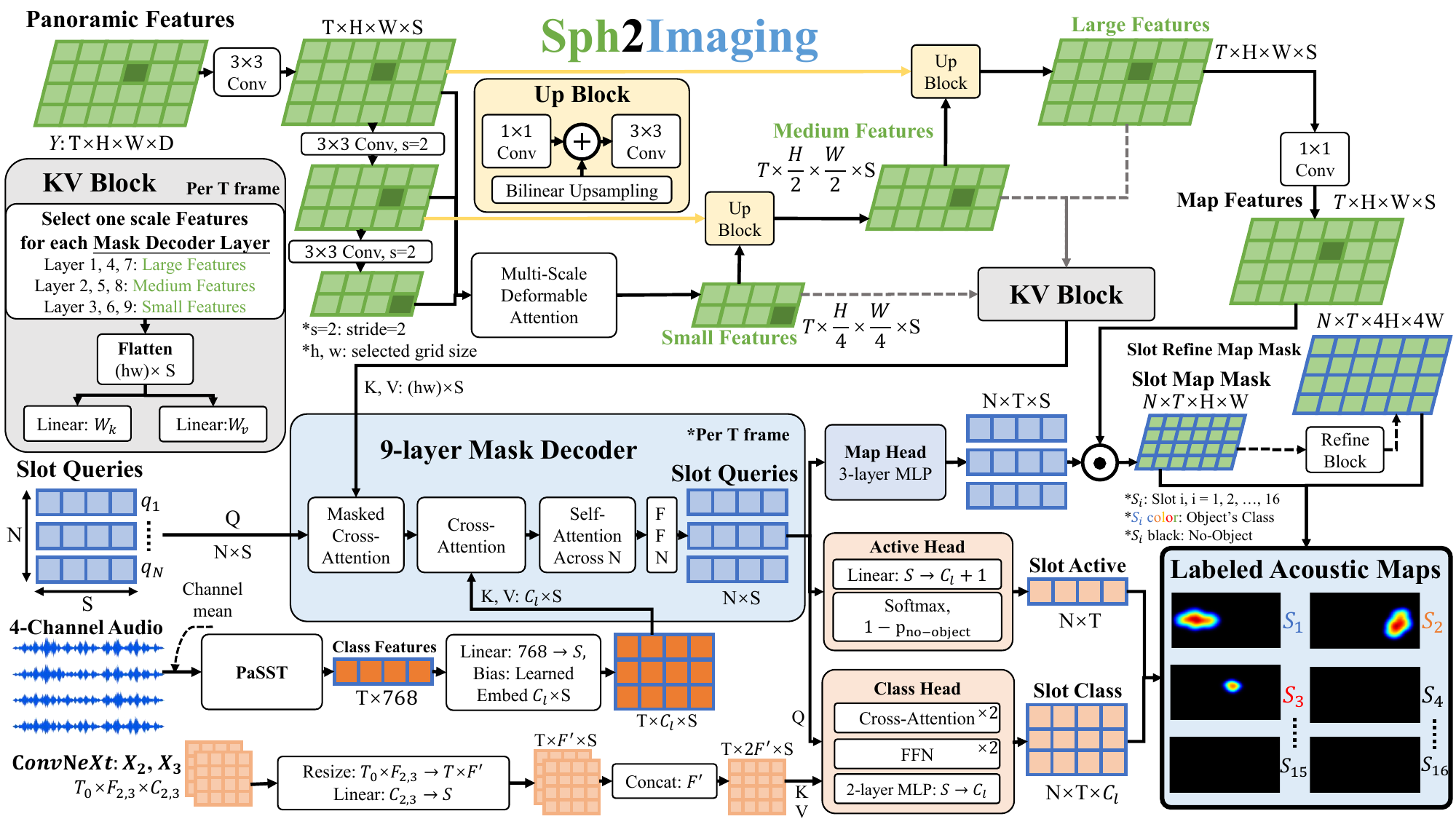}
\caption{Sph2Imaging updates slot queries using multi-scale spatial features and PaSST features. The Map Head predicts a map for each slot; the Active Head and Class Head estimate activity and class, respectively. The Refine Block produces higher-resolution maps for the final output.}
\label{fig:sph2imaging}
\end{figure*}

This paper is organized as follows.
Section 2 presents the SAID model.
Section 3 describes the data generation and training pipeline.
Section 4 reports the experimental results, and Section 5 concludes the paper.

\par\noindent\usebox{\saidoverview}\par

\section{Method}
\label{sec:method}
SAID converts four-channel audio into labeled acoustic maps (Fig.~\ref{fig:said}). Audio2Sph, the encoder, arranges audio features by direction; Sph2Imaging, the decoder, predicts a map, activity score, and class for each candidate source. Batch dimensions are omitted below.

\subsection{Audio2Sph: SAID's Encoder}
\label{sec:audio2sph}
At 48 kHz, a short-time Fourier transform (STFT) with a 2,048-sample Hann window, 2,048-point FFT, and 480-sample hop produces magnitude and phase at 100 fps (Fig.~\ref{fig:audio2sph}). An input convolution feeds concatenated multichannel log-magnitude and phase sine/cosine into four ConvNeXt stages~\cite{liu2022convnext}. Inter-stage kernels and strides are $(2,2)$, $(1,2)$, and $(1,2)$ in time-frequency order. The last three stages produce $X_i\in\mathbb R^{T_0\times F_i\times C_i}$, where $i=1,2,3$ indexes scales and $T_0,F_i,C_i$ count feature frames, frequency positions, and channels. For a two-second input, temporal padding and downsampling give $T_0=101$ feature frames.

Spherical Cross-Attention converts frequency features into direction features. A learned query $Q_{hw}\in\mathbb R^{1\times D}$ at row $h$ and column $w$ represents a direction on an $H\times W$ grid, with $H=45$ elevation rows, $W=90$ azimuth columns, and $D=16$. The three ConvNeXt outputs $X_1,X_2,X_3$ have different numbers of frequency positions and enter separate Spherical Cross-Attention blocks. These blocks share the same randomly initialized grid queries across all frames. For one ConvNeXt output, frame, and attention head, linear projections of the normalized query $Q_{hw}$ and normalized features $X\in\mathbb R^{F\times C}$ give $Q\in\mathbb R^{1\times d}$ and $K,V\in\mathbb R^{F\times d}$, where $d$ is the head width. Frequency weights $A\in\mathbb R^{1\times F}$ combine the values into $O\in\mathbb R^{1\times d}$~\cite{vaswani2017attention}:
\begin{equation}
A=\operatorname{softmax}_{F}\!\left(\frac{QK^{\mathsf T}}{\sqrt{d}}\right),\qquad O=AV.
\label{eq:spherical-cross-attention}
\end{equation}
Head outputs are concatenated, projected to $D$ channels, and added to $Q_{hw}$. A residual feed-forward network (FFN) produces the direction feature $Y_{hw}\in\mathbb R^{1\times D}$. Repeating across directions and frames produces $Y_i\in\mathbb R^{T_0\times H\times W\times D}$. Adding the three outputs at corresponding positions gives $Y_0=Y_1+Y_2+Y_3$. Time Alignment averages adaptive average- and max-pooled versions of $Y_0$ along time, producing Panoramic Features $Y\in\mathbb R^{T\times H\times W\times D}$ at 10 fps, where $T$ counts output frames.

During pretraining, $Y_0$ instead enters the Panoramic Decoder before Time Alignment. Three convolutional blocks with two intervening twofold bilinear upsampling steps, followed by a single-channel convolution and sigmoid, predict a $180\times360$ class-agnostic map of all active sources per frame. Temporal upsampling restores the STFT frame count. This supervision teaches Audio2Sph to locate sound energy; the Panoramic Decoder is then removed.

\subsection{Sph2Imaging: SAID's Decoder}
\label{sec:sph2imaging}
For each frame, a convolution projects $Y$ to $S=256$ channels; two stride-two convolutions successively halve the spatial dimensions, rounding up, to create coarser grids (Fig.~\ref{fig:sph2imaging}). Multi-Scale Deformable Attention~\cite{zhu2021deformable} uses grid features with positional information as queries to predict sampling locations and weights across all three grids. Weighted samples update the features; the coarsest output becomes Small Features. Two successive Up Blocks recover Medium and Large Features: each bilinearly upsamples to the corresponding finer grid's size, adds the corresponding finer convolutional grid after a $1\times1$ projection, and applies a $3\times3$ convolution~\cite{lin2017fpn}. A $1\times1$ convolution converts Large Features into Map Features $M\in\mathbb R^{T\times H\times W\times S}$.

PaSST~\cite{koutini2022passt} is initialized from the official AudioSet-pretrained PaSST-S checkpoint~\cite{passt_audioset_checkpoint}, further trained on DCASE training data, and then frozen. Pooling PaSST patch features over frequency and resizing time provide Class Features of shape $T\times768$. Projecting each frame to $S$ channels and adding $C_l=13$ learned class embeddings produces $T\times C_l\times S$ features, supplying $C_l$ keys and values per frame.

Sph2Imaging initializes $N=16$ learned Slot Queries of width $S$ per frame, each representing a candidate source rather than a fixed direction or class. The Map Head, a three-layer multilayer perceptron (MLP), maps query $q_n\in\mathbb R^S$ to $e_n\in\mathbb R^S$ for slot $n$. Within one frame, dot products with Map Features $M_{hw}\in\mathbb R^S$ and sigmoid $\sigma$ produce a continuous $45\times90$ Slot Map Mask~\cite{cheng2021maskformer}:
\begin{equation}
\mathrm{Mask}_n(h,w)=\sigma\!\left(e_n^{\mathsf T}M_{hw}\right).
\label{eq:slot-map-mask}
\end{equation}

The nine-layer Mask Decoder updates these queries independently per frame~\cite{cheng2022mask2former}. At each layer, the KV Block selects Large, Medium, then Small Features cyclically, flattens the selected grid, and projects the vectors into keys and values. Resizing the preceding map's dot-product values to this grid and thresholding their sigmoid at 0.5 defines a binary attention region. Masked Cross-Attention reads spatial features within this region; an empty region permits full-grid attention. Cross-Attention then reads the PaSST-derived features, followed by Self-Attention across slots and an FFN. Queries retain shape $N\times S$. The Map Head predicts maps before the first layer and after every update, supplying the next layer's attention region during both training and inference.

The final Slot Queries feed the Active and Class Heads. The linear Active Head predicts $C_l+1$ scores, including no-object; Slot Active is one minus the no-object softmax probability. For the Class Head, ConvNeXt features $X_2,X_3$ are each resized to $T\times F'$ ($F'=64$), projected to $S$ channels, and concatenated along frequency into $T\times2F'\times S$. Per frame, these features supply keys and values for two cross-attention and FFN layers, with the final Slot Queries as queries. The output is scaled by a learned weight and added to the input queries; a two-layer MLP gives $C_l$ Slot Class scores, which softmax converts to class probabilities.

The $45\times90$ Slot Map Mask guides decoder updates; the Refine Block produces $180\times360$ maps for training supervision and inference. Two stride-two transposed convolutions upsample Map Features, while a separate MLP projects the final Slot Queries to the same channel width. Dot products followed by sigmoid produce the refined maps.

Each refined map receives the class with the highest Slot Class probability. Confidence is the product of three factors: Slot Active, the highest Slot Class probability, and the mean refined-map value over pixels above 0.5. If no pixels exceed 0.5, confidence is zero. Class labels and confidence leave map values unchanged.

\section{Data and Training}
\label{sec:training}
\subsection{Online Scene Generation}
We simulate two-second four-channel recordings online with the image source method (ISM)~\cite{allen1979image} in Pyroomacoustics~\cite{scheibler2018pyroomacoustics}, sampling rooms and source positions and summing source signals at the microphones. Pretraining uses 0--4 VCTK~\cite{veaux2019vctk} sources per scene, room widths and lengths of 2--10 m, heights of 2--4 m, reflection orders up to five, and wall absorption sampled uniformly from $[0,0.5]$.

For class-labeled training, we construct SourceBank by collecting 122,359 clips (about 95.4 hours) from VCTK, MUSDB18-HQ~\cite{rafii2019musdb}, FSD50K~\cite{fonseca2022fsd50k}, and verified FSDKaggle2018~\cite{fonseca2018freesound} clips, mapped to 13 DCASE classes. Each scene contains 1--6 sources. Since ISM models point sources, we distribute multiple emitting points across a region and drive all points with the same clip to simulate an extended source. Source positions, extents, and activity determine each source's target map; its class label comes from the selected clip.

\subsection{Staged Training}
First, we pretrain Audio2Sph with the Panoramic Decoder, bypassing Time Alignment to predict $180\times360$ maps at 100 fps. At each frame, spherical Gaussians centered on valid, active sources have angular standard deviation $4^\circ$. Their pointwise maximum forms the target map; frames without active sources have an all-zero target. For sigmoid predictions $P_{\mathrm{pre}}$ and targets $Y_{\mathrm{pre}}$, unweighted binary cross-entropy (BCE) averages over batch, time, and pixels:
\begin{equation}
\mathcal L_{\mathrm{pretrain}}=\operatorname{BCE}_{\mathrm{mean}}(P_{\mathrm{pre}},Y_{\mathrm{pre}}).
\label{eq:pretraining-loss}
\end{equation}
Pretraining uses 2.5 million steps with batch size one.

Next, we retain the pretrained encoder, replace the Panoramic Decoder with Sph2Imaging, and train SAID on SourceBank scenes at 10 fps. Predictions from the initial Slot Queries and each of the nine Mask Decoder layers form ten groups~\cite{cheng2022mask2former}. At each frame, Hungarian matching pairs slots with sources independently for each group, using map and class agreement~\cite{cheng2021maskformer,carion2020detr}.

The Map Head receives BCE and Dice supervision on matched $45\times90$ maps at uncertainty-guided and random sample positions. Summing across the ten groups gives $\mathcal L_{\mathrm{BCE}}$ and $\mathcal L_{\mathrm{Dice}}$. The $180\times360$ refined maps use the final prediction group's matching. Their loss $\mathcal L_{\mathrm{refine}}$ combines full-map BCE and Dice (superscript $\mathrm{refine}$) with $\mathcal L_{\mathrm{corr}}$ and $\mathcal L_{\mathrm{sIoU}}$: one minus mean Pearson correlation and soft intersection-over-union across matched maps, respectively.

The Active Head uses a linear layer to efficiently predict class probabilities for matching during training. Supervising the Active Head encourages the Slot Queries to encode class information that the Class Head can decode for the final prediction. Cross-entropy (CE) targets the source class for matched slots and no-object for unmatched slots, with weights 1 and 0.1, respectively. Summing weighted mean CE over the ten groups gives $\mathcal L_{\mathrm{CE}}$.

The Class Head determines the output class. Using the final matching, matched slots receive focal loss $\mathcal L_{\mathrm{focal}}$~\cite{lin2017focal}, with $\gamma=2$ and inverse-square-root class-frequency weights normalized to mean one and clipped to $[0.25,4]$.

The complete objective is
\begin{equation}
\begin{aligned}
\mathcal L_{\mathrm{said}}={}&2\mathcal L_{\mathrm{CE}}+5\mathcal L_{\mathrm{BCE}}+5\mathcal L_{\mathrm{Dice}}\\
&+\mathcal L_{\mathrm{refine}}+\mathcal L_{\mathrm{focal}},
\end{aligned}
\label{eq:complete-training-loss}
\end{equation}
\begin{equation}
\begin{aligned}
\mathcal L_{\mathrm{refine}}={}&\mathcal L_{\mathrm{BCE}}^{\mathrm{refine}}+\mathcal L_{\mathrm{Dice}}^{\mathrm{refine}}\\
&+0.5(\mathcal L_{\mathrm{corr}}+\mathcal L_{\mathrm{sIoU}}).
\end{aligned}
\label{eq:refinement-loss}
\end{equation}
Finally, we fine-tune on DCASE recordings with annotated maps and classes using the same objective~\cite{roman2026stairs26,shimada2023starss23}. Augmentation gives four views per segment by selecting four-channel microphone configurations for azimuth rotations of $0^\circ$, $90^\circ$, $180^\circ$, and $270^\circ$ and shifting target maps horizontally to match.

\section{Experiments}
\label{sec:experiments}

\subsection{Evaluation and Results}

We evaluate prediction compression on the development test set and report challenge results on the official DCASE2026 Task~3 Track~A evaluation set. The evaluation set contains 79 recordings totaling approximately 3.5 hours and covers 13 sound classes. Systems predict labeled acoustic maps at 10 frames per second using four-channel audio without video~\cite{dcase2026task3}.

Macro mean average precision (Macro mAP) evaluates class predictions and source-region overlap across multiple overlap thresholds, penalizing missed sources and false detections. Macro Pearson~$r$ measures energy-pattern correlation between spatially matched predictions and references of the same class. Both metrics are averaged across classes, and the official ranking uses the sum of their ranks~\cite{dcase2026task3}.

To meet the 20 MB limit per recording, we store refined maps as sparse points~\cite{dcase2026task3}. Each grid cell retains its strongest pixel at least 10\% of the map peak. Grid spacing is 2 pixels for confidence scores of at least 0.20 and 6 otherwise. Lower-confidence maps receive support points offset 2 pixels from low-energy boundary pixels and assigned 12\% of peak energy. All detections are retained. The official evaluator reconstructs maps using Gaussian smoothing~\cite{dcase2026task3}.

Table~\ref{tab:compression} shows a 97.81\% reduction in total JSON storage across 78 development-test recordings, with every file below 20 MB. Macro mAP decreases by 0.0024, while Macro Pearson~$r$ increases by 0.0219.

\begin{table}[!ht]
\caption{Prediction compression on the full-recording development test set (78 recordings). JSON sizes are in decimal MB per recording.}
\label{tab:compression}
\centering
\setlength{\tabcolsep}{2pt}
\begin{tabular*}{\columnwidth}{@{}l@{\extracolsep{\fill}}rrrr@{}}
\toprule
Prediction & Macro mAP & Macro Pearson $r$ & Max. JSON & Avg. JSON \\
\midrule
Original & 0.1202 & 0.4268 & 941.50 & 369.58 \\
Compressed & 0.1177 & 0.4487 & 18.70 & 8.08 \\
\bottomrule
\end{tabular*}
\par\vspace{-3pt}
\end{table}

Table~\ref{tab:official} compares the best-ranked submission from each Track~A team. Our challenge submission ranks first with 0.1080 Macro mAP and 0.3962 Macro Pearson~$r$, achieving the highest scores in both columns.

\begin{table}[!ht]
\vspace{-6pt}
\caption{Official DCASE2026 Task~3 Track~A results, using each team's best-ranked submission~\cite{dcase2026task3,dcase2026task3results}. Both metrics are macro-averaged; higher scores are better.}
\label{tab:official}
\centering
\begin{tabular}{clcc}
\toprule
Rank & Team & Macro mAP & Macro Pearson $r$ \\
\midrule
1 & CUHK (SAID) & \textbf{0.1080} & \textbf{0.3962} \\
2 & Medisensing & 0.0662 & 0.2765 \\
3 & Wuhan University & 0.0091 & 0.3185 \\
4 & Samsung Electronics & 0.0215 & 0.1394 \\
5 & Deka & 0.0001 & 0.0274 \\
\bottomrule
\end{tabular}
\par\vspace{-8pt}
\end{table}

\subsection{Ablation Studies}

Table~\ref{tab:ablation} compares SAID (PaSST) and three variants on 78 development-test recordings. First, SAID (AudioMAE) replaces PaSST with AudioMAE~\cite{huang2022audiomae,gaunernst_audiomae_checkpoint}, while SAID (None) removes the PaSST branch; both retain the Class Head. Second, \emph{w/o Audio2Sph pretraining} jointly trains Audio2Sph and Sph2Imaging without first pretraining Audio2Sph.

We compute two additional metrics from the same predictions. Mask AP measures class-agnostic source-region detection using the official average precision (AP) calculation with all classes merged. AP summarizes precision across recall levels. Macro Class-F1 evaluates classification only for spatially matched sources. Per frame, Hungarian matching pairs predicted and reference map peaks by angular distance, without class labels. Pairs within $20^{\circ}$ are pooled across recordings to compute per-class F1, the harmonic mean of precision and recall. F1 is averaged over all classes present in the reference set, assigning zero when undefined; unmatched sources are excluded.

\begin{table}[!ht]
\vspace{-6pt}
\caption{Ablations on the development test set. Mask AP is class-agnostic; Macro Class-F1 uses only spatial matches within $20^{\circ}$. Higher scores are better.}
\label{tab:ablation}
\centering
\setlength{\tabcolsep}{2pt}
\begin{tabular*}{\columnwidth}{@{}l@{\extracolsep{\fill}}cccc@{}}
\toprule
Variant & \shortstack{Macro\\mAP} & \shortstack{Macro\\Pearson $r$} & \shortstack{Mask\\AP} & \shortstack{Macro\\Class-F1} \\
\midrule
SAID (PaSST) & 0.1202 & 0.4268 & 0.2380 & 0.3885 \\
SAID (AudioMAE) & 0.1134 & 0.4307 & 0.2287 & 0.3951 \\
SAID (None) & 0.0260 & 0.4224 & 0.2293 & 0.1347 \\
w/o Audio2Sph pretraining & 0.0000 & 0.0510 & 0.0000 & 0.3785 \\
\bottomrule
\end{tabular*}
\par\vspace{-8pt}
\end{table}

\section{Conclusion}
\label{sec:conclusion}

We presented SAID for predicting a separate labeled acoustic map for each active sound source from four-channel audio. Audio2Sph learns panoramic features arranged by direction, and Sph2Imaging decodes these features into source regions, energy, and classes. Our training pipeline first pretrains Audio2Sph without class labels using online simulation, then trains SAID to predict individual source maps and classes before fine-tuning on real recordings. Our challenge submission ranks first on the official DCASE2026 Task~3 Track~A evaluation set, achieving 0.1080 Macro mAP and 0.3962 Macro Pearson~$r$. The model and data-generation pipeline provide a starting point for further research on audio-only semantic acoustic imaging.

\section*{Acknowledgement}
This work was supported by the Innovation and Technology Fund (ITF), Hong Kong, under Project ITS/301/24.

\bibliographystyle{IEEEtran}
\bibliography{references}
\end{document}